# DK And:
## Reclassification as EW Binary from CCD Observations

Hambsch, F.-J.[1] and Husar, D.[2,3]

[1] present adress: Dr. Franz-Josef Hambsch, Oude Bleken 12, 2400 Mol-Millegem (Belgium)
[2] Himmelsmoor Private Observatory Hamburg, Himmelsmoor 18, 22397 Hamburg (Germany)
[3] present adress: Dr. Dieter Husar, c/o EAT SA, Rue du Séminaire 20A, 5000 Namur (Belgium);
  email : husar.d (at) gmx.de

also including observations made by
Poschinger, K.v. [4] and Walter, F.[5]

[4] present adress: Konstantin von Poschinger, Hammerichstr. 5, 22605 Hamburg (Germany)
[5] present adress: Frank Walter, Denninger Str. 217, 81927 München (Germany)

***Abstract:***

*This paper describes the reclassification of DK And, formerly classified as a RRc type star, as EW binary. 1599 CCD unfiltered and filtered (V and R band) observations between 1999 and 2005 show, that the star is actually an eclipsing binary star with a period of **P = 0.4892224 ± 0.0000002 [d]** with epoch **E₀ = 2451435.4353 ± 0.0010** (if all historic data were taken into account). From our new observations 12 timings of minimum light are given.*

## 1) Introduction

The variability of DK And was first analysed by Götz, Huth and Hoffmeister in 1957 [1]. They derived a period of $P = 0.243655$ [d] with the epoch $E_0 = 2429130.407$ and an amplitude of the light variation of $\Delta m = 0.5$ mag. These elements of reference [1] are also the ones found in the GCVS. In the original work [1] only a rapid magnitude change of the star is discussed, but the data were not precise enough to determine the nature of the star, whether it belongs to the group of RR Lyrae stars or the group of W Ursae Majoris stars. The rise time M-m which was mentioned to be 0.12 [d] is about half of the period. The light curve was already recognized to be quite symmetrical.

Kemper in 1982 [2] did not eliminate DK And in his detailed spectroscopic study of RRc Lyrae stars as being an eclipsing binary variable although he did this for other stars.

In 1992 Ratcliff [3] attributes this star in his list of short-periodic variables as 'RR**:**' without giving any reason for his classification.

Hence, none of the mentioned literature indicates without doubts that DK And is not a RR Lyrae star.

The GEOS RR Lyrae database [4] contains 10 times of maxima covering the time from 1938 to 2005. Significant deviations of the calculated to the observed maximum ***(O-C)*** indicate already that in the GCVS wrong elements were given for DK And.

## 2) New CCD observations and reclassification

In our present analysis we have used new CCD observations from the years 1999 to 2005: the star has been observed by Husar (HSR) during 4 nights in the years 1999 to 2004, by Poschinger (PC) [5] during 5 nights between 2003 and 2005, by Walter (WTR) [6] in 4 nights between 2004 and 2005. Hambsch (HMB) observed the star in 2 nights in 2005 with V and R band filters. In total 1599 CCD measurements have been made covering a total of 2236 days.

As reference and/or comparison stars we used: GSC 3649-0879 (V = 12.7 mag; USNO A2.0: R = 13.0 mag); GSC 3649-1549 (V = 12.35 mag; USNO A2.0: R = 12.5 mag) and GSC 3645-1799 (V = 12.21 mag; USNO A2.0: R = 12.5 mag).

The amount of data has been reduced with the period search and analysis software PERANSO Version 2.1 [7]. Only little corrections of the relative magnitudes had to be made because of the different instrumentation to form a homogeneous light curve (see Fig. 1).

According to the highly symmetric form of the reduced light curve it is already very probable that DK And is an EW binary. The amplitude of $\Delta m$ = 0.55 mag, which is also typical for EW binaries, does not contradict this classification. After analysing all the data with a doubled period we could not find clear evidence for different minimum depth. Already with a preliminary period of $P$ = 0.489223 [d] we find a light curve (Fig. 1) which favours the classification of DK And as an EW binary.

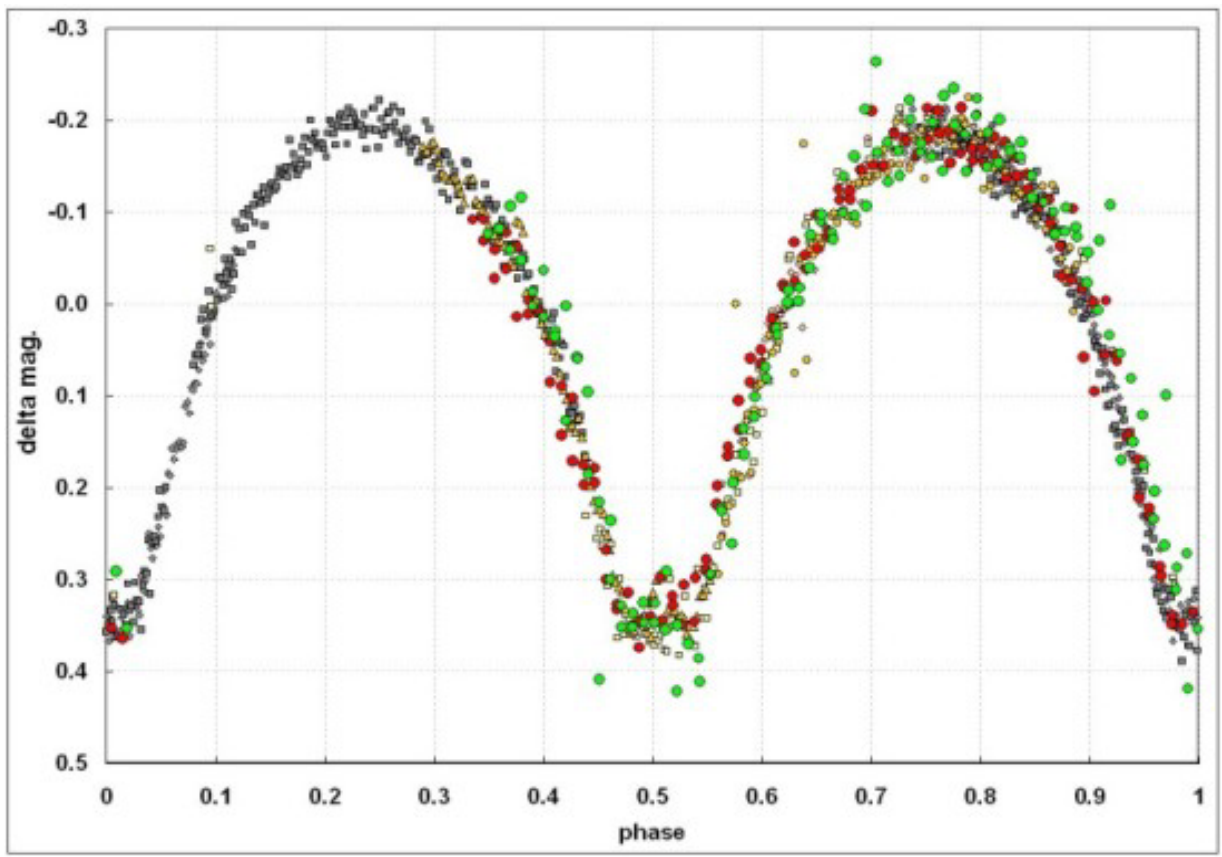

**Fig. 1:** reduced light curve of DK And for the mentioned CCD observations between 1999 and 2005
(only observations from HMB, HSR and WTR: HMB V-filtered = green and R-filtered = red circles, HSR = grey squares and diamonds, WTR = yellowish circles and squares)
calculated with our preliminary elements: $E_0$ = 2451435.434 and $P$ = 0.489223 [d]

The strongest argument for the reclassification of DK And however comes from the fact that the observations with V and R broad band filters (Bessel type) do not show any significant indication for a colour change along with the phase as can be seen from Fig. 2.

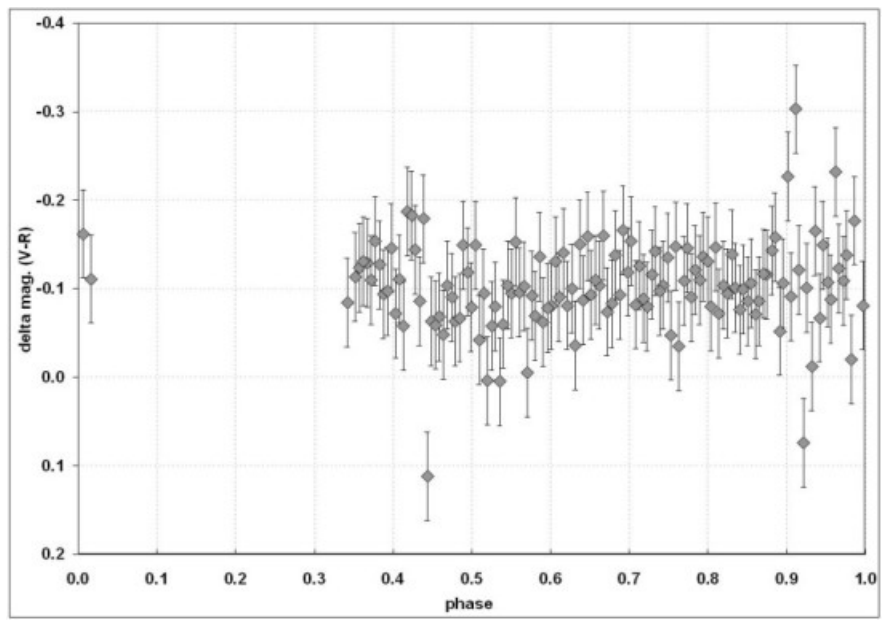

**Fig. 2:** phased V-R light curve of DK And from the filtered CCD observations by HMB
calculated with our new elements: $E_0$ = 2451435.4353 and $P$ = 0.4892224 [d]

### 3) Analysis of ROTSE-I data

Further on we analysed the data from the ROTSE-I project [8] made publicly available by the NSVS in 2004 [9]. Using our new elements the reduced light curve out of 132 good data points is shown in Fig.3. An amplitude of about $Δm$ = 0.6 mag agrees with our own results. There seems to be a slight shift as far as the determination of the epoch is concerned. However we attribute this being due to the large photometric errors of the ROTSE-I data (median error: ± 0.05 mag; magnitude scatter: ± 0.128 mag according to NSVS information given for this star in Ref. [9]). The difference between primary minimum and secondary minimum seems to be more evident but again the poor photometric quality of the data would not have allowed a final conclusion only from these data.

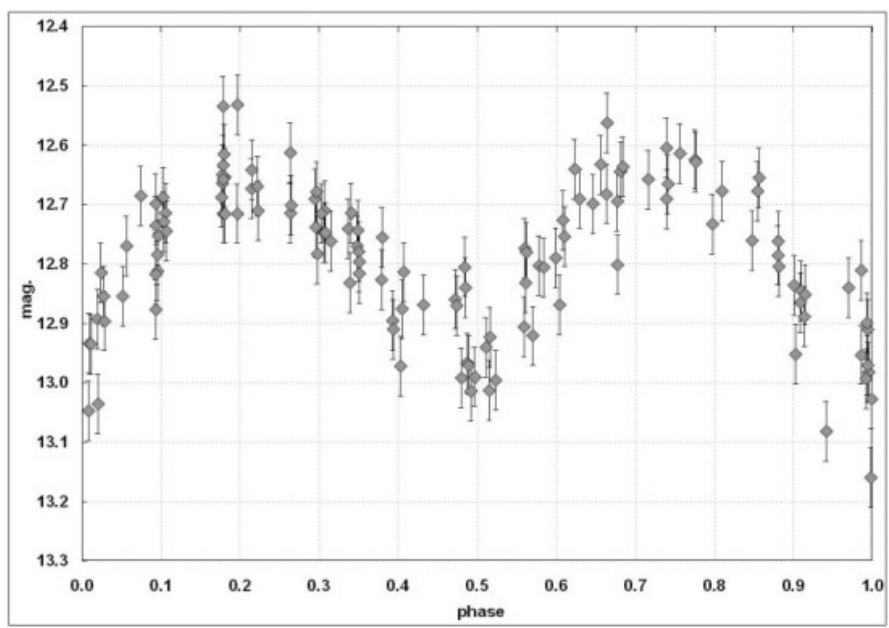

**Fig. 3:** reduced light curve of DK And based on ROTSE-I data [8] from NSVS [9]
- median errors of the individual data points were ± 0.05 mag (according to NSVS [9])
- phases were calculated with our new elements: $E_0$ = 2451435.4353 and $P$ = 0.4892224 [d]

## 4) New CCD minima and new elements for DK And

In order to obtain a more precise determination of the period we extracted all minima from the new CCD observations. All results from Poschinger, which were already published as maxima, we reanalysed in order to extract the minimum timings from the original data.

For the minimum timings we used the determination of the minimum of a polynomial and/or the Kwee van Woerden method which are now both implemented in the Peranso software [7]. The systematic errors due to the different timing methods are contained in the estimated errors for the maximum timings given in Table 1.

Taking only all these new CCD data we derived new elements for DK And from a least square fit:

$$\text{HJD(Max)} = 2451435.4317 + 0.4892238 \text{ [d]} \times E \qquad (1)$$
$$\pm\ 0.0004 \pm 0.0000002 \text{ [d]}$$

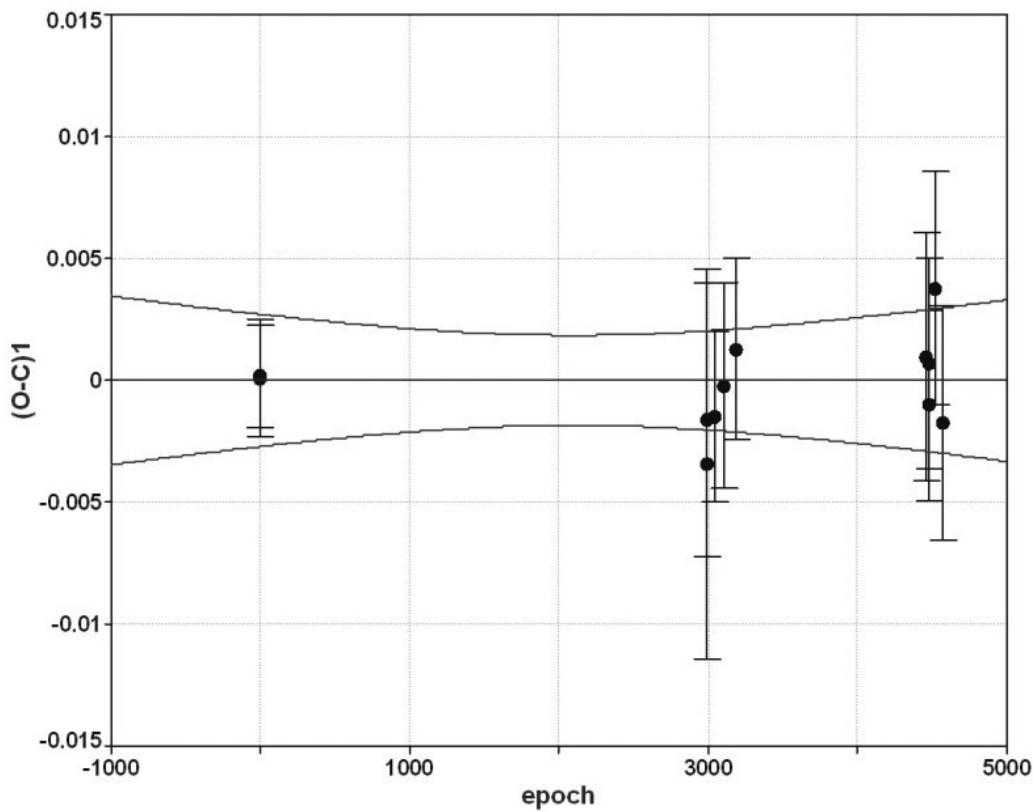

**Fig. 4:** *(O-C)$_1$* diagram of DK And
- only with CCD observations from HMB, HSR, PC and WTR
- *(O-C)$_1$* values are calculated with the elements (1): $E_0$ = 2451435.4317 and $P$ = 0.4892238 [d]
- the graph shows the 99.9% confidence intervals

## 5) Calculation of new elements for DK And from all available data

In order to improve the precision of the derived period $P$ an attempt to include the historic photographic observations (the first dating back to 1938) in the calculation was made in the following way: for all photographic maximum timings from Götz et al. [1] and for the CCD maximum of Birkner (as taken from the GEOS database [4]) we subtracted $P$/4 from the published times of maxima. All results are represented together in Table 1 of this paper.

The analysis of all available data together resulted in the following new elements for DK And:

$$\text{HJD(Max)} = 2451435.4353 + 0.4892224 \text{ [d]} \times E \qquad (2)$$
$$\pm 0.0010 \pm 0.0000002 \text{ [d]}$$

A significant difference can be seen between the elements **(1)** from the new CCD observations and the elements above **(2)** which incorporate all available historical data for DK And. This is however not so surprising as the time span of the new CCD observations is short compared with the time span of all historical data. On the other hand the difference of the periods $\Delta P(1-2) = 1.4 \times 10^{-6}$ is more than 4 times higher compared with the calculated error ($\pm 0.3 \times 10^{-6}$) of the difference of the periods. This may either result from underestimation of the errors of the minimum timings or it may indicate a secular period change. We do not want to overemphasize the difference, as we know about the difficulties to estimate the timing errors correctly, but we would like to encourage further observations of DK And in order to improve the elements or observe eventual changes.

| epoch | type of MIN | HJD (MIN) | ± * | (O-C)1 | (O-C)2 | rem./obs. |
|---|---|---|---|---|---|---|
| -45593 | calc. prim. | 2429130.29 | 0.03 | 0.0387 | -0.0287 | Götz |
| -45532 | calc. prim. | 2429160.23 | 0.03 | 0.1371 | 0.0697 | Götz |
| -44563 | calc. prim. | 2429634.19 | 0.03 | 0.0352 | -0.0308 | Götz |
| -43461 | calc. prim. | 2430173.37 | 0.03 | 0.0966 | 0.0321 | Götz |
| -35808 | calc. prim. | 2433917.37 | 0.03 | 0.0618 | 0.0081 | Götz |
| -35792 | calc. prim. | 2433925.19 | 0.03 | 0.0542 | 0.0005 | Götz |
| -2847.5 | calc. sec. | 2450042.376 | 0.002 | 0.0096 | 0.0020 | Birkner |
| 0 | prim. | 2451435.4319 | 0.0021 | 0.0002 | -0.0034 | HSR |
| 2 | prim. | 2451436.4102 | 0.0024 | 0.0001 | -0.0035 | HSR |
| 2990.5 | sec. | 2452898.4539 | 0.0056 | -0.0016 | -0.0010 | PC |
| 2997 | prim. | 2452901.6320 | 0.0080 | -0.0034 | -0.0028 | HSR |
| 3050 | prim. | 2452927.5628 | 0.0035 | -0.0015 | -0.0008 | PC |
| 3109 | prim. | 2452956.4283 | 0.0042 | -0.0002 | 0.0005 | PC |
| 3190.5 | sec. | 2452996.3016 | 0.0037 | 0.0013 | 0.0022 | PC |
| 4464.5 | sec. | 2453619.5723 | 0.0051 | 0.0010 | 0.0036 | PC |
| 4478.5 | sec. | 2453626.4212 | 0.0043 | 0.0007 | 0.0033 | HMB |
| 4480.5 | sec. | 2453627.3979 | 0.0039 | -0.0010 | 0.0016 | HMB |
| 4527.5 | sec. | 2453650.3963 | 0.0048 | 0.0038 | 0.0066 | WTR |
| 4570.5 | sec. | 2453671.4274 | 0.0048 | -0.0017 | 0.0011 | WTR |

**Table 1:** All available times of minima for DK And
in blue: photographic (calculated from the published maximum by subtracting P/4)
in green: CCD observation (calculated from the published maximum by subtracting P/4)
in black: new minimum timings from recent CCD observations
$(O-C)_1$ values are calculated with the elements **(1)** given in this paper
$(O-C)_2$ values are calculated with the elements **(2)** given in this paper
* estimated errors of minimum timings
rem./obs.: observers HMB = Hambsch, HSR = Husar, PC = Poschinger, WTR = Walter

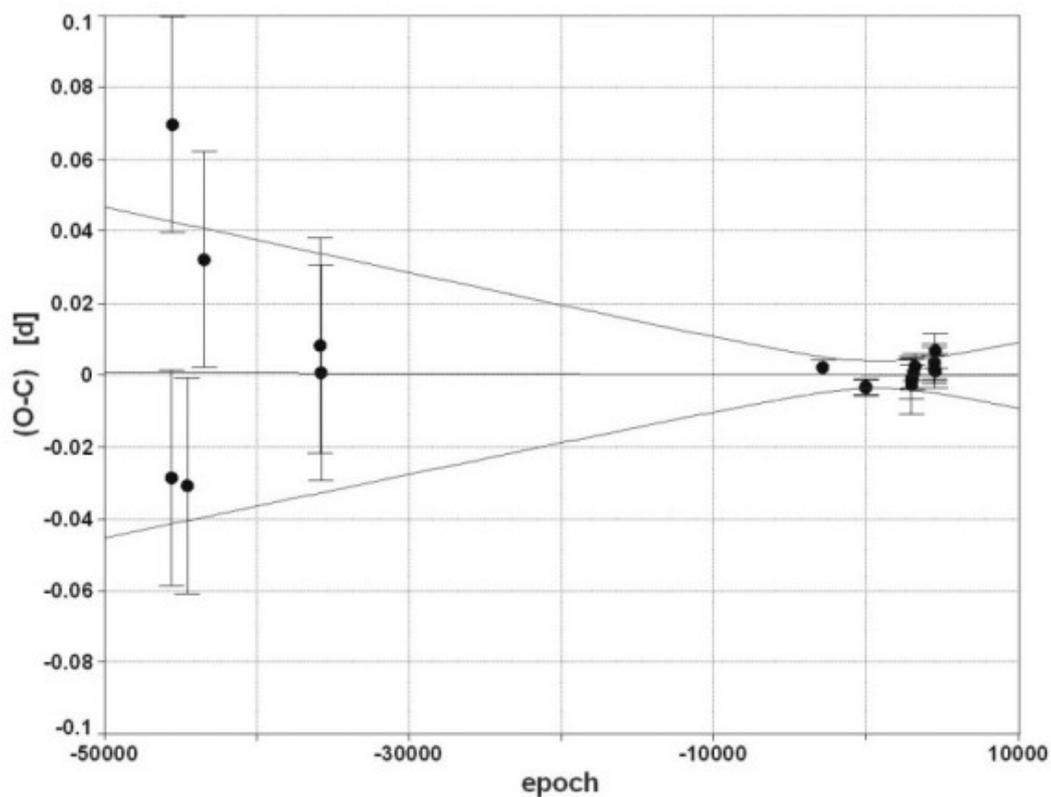

**Fig. 5:** *(O-C)₂* diagram of DK And
- all available data: (calculated) photographic minima and all CCD observations
- *(O-C)₂* values are calculated with the elements (2): $E_0$ = 2451435.4353 and $P$ = 0.4892224 [d]
- the graph shows the 99.9% confidence intervals

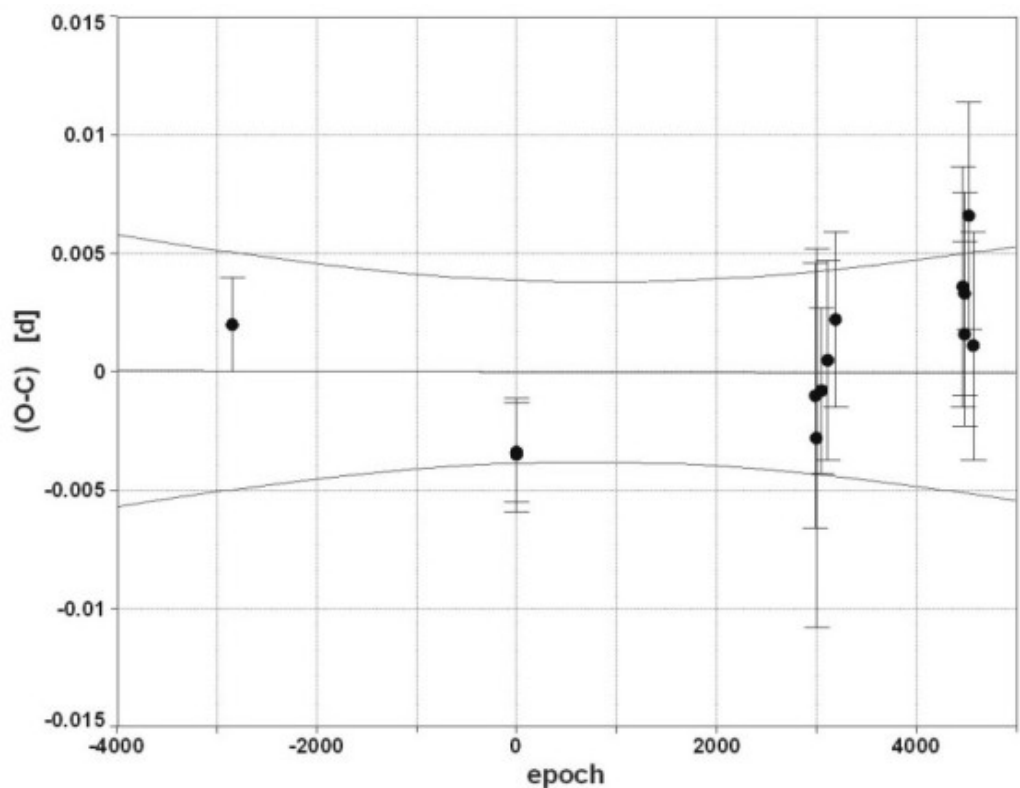

**Fig. 6:** zooming into the right part of Fig. 5: *(O-C)₂* diagram of DK And
- between epoch -4000 and epoch 5000 = only the CCD observations
- *(O-C)₂* values are calculated with the elements (2): $E_0$ = 2451435.4353 and $P$ = 0.4892224 [d]
- the graph shows the 99.9% confidence intervals

## 6) Acknowledgements